# Critical Exponents of the Superconducting Phase Transition

Michael Kiometzis, Hagen Kleinert, and Adriaan M. J. Schakel

Institut für Theoretische Physik

Freie Universität Berlin

Arnimallee 14, 14195 Berlin

## Abstract

We study the critical exponents of the superconducting phase transition in the context of renormalization group theory starting from a dual formulation of the Ginzburg-Landau theory. The dual formulation describes a loop gas of Abrikosov flux tubes which proliferate when the critical temperature is approached from below. In contrast to the Ginzburg-Landau theory, it has a spontaneously broken global symmetry and possesses an infrared stable fixed point. The exponents coincide with those of a superfluid with reversed temperature axis.

7440, 6470

In a recent paper we have pointed out the advantage of using a dual formulation of the Ginzburg-Landau theory to describe the superconducting phase transition in three space dimensions ($D = 3$) [1]. The theory is a continuum version of a lattice model put forward some time ago by one of the present authors to study this transition as well as the tricritical point [2]. Recently, Kovner, Kurzepa, and Rosenstein [3] argued that the continuum model could account for all fixed points which one expects in the Ginzburg-Landau theory, not just the infrared fixed point. It focusses on the Abrikosov flux tubes in the theory, which carry magnetic flux, rather than the electrically charged Cooper pairs. Whereas the Ginzburg-Landau theory features a local U(1) gauge symmetry, the dual theory involves only a *global* U(1) symmetry. To understand the relevance of this difference, we recall that a basic element of Landau's theory of continuous phase transitions is the presence of an order parameter. When this parameter develops an expectation value, which happens at the critical temperature, some global symmetry is broken. For a local symmetry no order parameter exists in the sense of Landau, implying that such a symmetry can never be broken [4]. A gauge theory seems consequently not appropriate to describe a second-order phase transition. This may explain the fact why an infrared stable fixed point was never been found within the Ginzburg-Landau formulation of the superconducting phase transition [5,6], although it is accepted that the transition is of second order in the type II regime, and thus should possess such a point.

Following a suggestion by Helfrich and Müller [7], Dasgupta and Halperin [8] simulated a lattice model to study the problem. With help of a dual transformation [9,10] they found for a whole range of parameters a second order phase transition with superfluid exponents. One of us [11] succeeded in mapping, via a duality transformation, the lattice Ginzburg-Landau theory onto a $|\psi|^4$ theory, and was able to obtain the complete phase diagram analytically. In particular, he predicted the existence and located the position of a tricritical point.

The dual formulation of the Ginzburg-Landau theory has a global U(1) symmetry, which is spontaneously broken when the system crosses from the superconducting into the normal phase. That is, in contrast to the Ginzburg-Landau theory itself, the dual formulation pos-sesses an order parameter to describe the phase transition. To understand the symmetry involved, we interpret the three-dimensional theory as one in two space and one time dimensions. This allows us to make use of results obtained by Kovner and Rosenstein in part in collaboration with Eliezer for scalar QED in 2+1 dimensions [12]. They pointed out that the relevant symmetry for the Coulomb-Higgs phase transition in scalar QED, which corresponds to the superconducting phase transition in the Ginzburg-Landau theory, is the global U(1) symmetry generated by the magnetic flux $\Phi = \int d^2x \tilde{F}_0$. Here, $\tilde{F}_0$ is the zero component of the dual field strength

$$\tilde{F}_\mu = \epsilon_{\mu\nu\lambda}\partial_\nu A_\lambda, \tag{1}$$

with $\epsilon_{\mu\nu\lambda}$ ($\mu,\nu,\lambda = 0,1,2$) being the antisymmetric Levi-Civita symbol and $A_\mu$ the electromagnetic gauge potential. This symmetry is referred to as magnetic flux symmetry. In the phase where the photon is massless, corresponding to the high-temperature phase of a superconductor, the magnetic flux symmetry is broken. The resulting Goldstone particle is the photon. (It should be noted that in 2+1 dimensions the photon has only one transverse direction and thus only one degree of freedom, just like a scalar particle.) In the Higgs, or superconducting phase, the magnetic flux symmetry is unbroken.

The current associated with the flux symmetry is the dual field strength $\tilde{F}_\mu$. When written in terms of the gauge potential—as in Eq. (1)—this current is a topological current which is trivially conserved due to the presence of the $\epsilon$-symbol. In the dual description, this symmetry becomes explicit and the associated current is a genuine Noether current.

The dual theory of a $3D$ superconductor, which is defined by the bare Hamiltonian [1–3]

$$H_{\psi,0} = |(\partial_i - ig_0 h_{0,i})\psi_0|^2 + m_0^2|\psi_0|^2 + u_0|\psi_0|^4 + \frac{1}{2}(\nabla \times \mathbf{h}_0)^2 + \frac{1}{2}m_{A,0}^2\mathbf{h}_0^2, \tag{2}$$

describes a complex scalar field $\psi_0$ with mass parameter $m_0$ and a contact interaction $u_0|\psi_0|^4$. It is minimally coupled to a *massive* vector field $h_{0,i}$, $i = 1,2,3$, with a coupling constant $g_0 = (2\pi/e_0)m_{A,0}$, where $e_0$ is the electric charge of the scalar field in the Ginzburg-Landau theory. The vector field has the same mass $m_{A,0}$ as the electromagnetic gauge field in the superconducting phase of the Ginzburg-Landau theory. Since $h_{0,i}$ is massive, the theory lacks local gauge symmetry. We have scaled the coupling constants and the fields in such a way that no explicit temperature dependence appears in the partition function $Z$, which is described by the functional integral

$$Z = \int Dh_{0,i} D\psi_0^* D\psi_0 \exp\left(-\int d^3x H_{\psi,0}\right), \qquad (3)$$

where the field $\mathbf{h}_0$ is treated as an independent fluctuating field.

To acquire an intuitive understanding of the dual theory, we picture an Abrikosov flux tube as a line-like object carrying one unit of magnetic flux. The scalar field $\psi_0$ gives a field theoretic description of a loop gas of these objects, i.e., of closed magnetic flux tubes. Their mass $m_0$ physically represents the energy per unit length of a flux tube. This construct applies of course only to type II superconductors; for type I superconductors no stable flux tubes exist. The coupling to the vector field $h_{0,i}$ accounts for the fact that the flux loops carry magnetic flux. In fact, $\mathbf{h}_0$ represents the microscopic, or local field.

In the Ginzburg-Landau theory, the local field becomes singular at the lines representing the loops. In the spirit of Dirac [13], these singularities are subtracted by introducing a plastic tensor $h_i^{\text{P}}$ [14], which is a singular field that yields a $\delta$-function along the flux loops. The physical local field is now given by $(\nabla \times \mathbf{A})_i - h_i^{\text{P}}$, which is everywhere regular. In the dual description, this combination is represented by $h_{0,i}$; the plastic tensor $h_i^{\text{P}}$ itself is represented by the Noether current $j_i = \psi^* \overleftrightarrow{\partial}_i \psi - 2igh_i\psi^*\psi$.

Although at first sight a theory like (2) with a massive vector field looks perturbatively nonrenormalizable in four dimensions ($D = 4$) [15], a closer inspection reveals that it is renormalizable [16]. We therefore can apply usual perturbation theory to calculate the critical exponents. The derivation of the dual theory (2) from the Ginzburg-Landau model hinged on the fact that the number of dimensions is three, so that the dual object $\epsilon_{ijk}\partial_j A_k$ is a vector. In other words, the dual theory describes the superconducting phase transition only in three dimensions. For this reason, we study the model in fixed ($D = 3$) dimension, and not in $D = 4 - \epsilon$ dimensions as is often done. The fixed-dimension approach to critical phenomena was introduced by Parisi, who applied it to a pure $|\psi|^4$ theory [17]. The method makes explicitly use of the fact that near the critical point the system has only one relevant length scale, viz. the correlation length which diverges at this point. This length is used to convert dimensionful coupling constants into dimensionless ones.

In the present setting, the field $\psi$ develops an expectation value when the critical temperature is approached from below, indicating the occurence of a flux tube condensate. So the relevant scale is the (renormalized) inverse mass $m^{-1}$. (The bare mass vanishes as $m_0^2 \sim T_c - T$ at the critical temperature $T_c$). Of course, since the mass of the vector field is identified with the photon mass it also vanishes at the transition temperature. As we know from the Ginzburg-Landau theory, the bare mass vanishes as $m_{A,0}^2 \sim T_c - T$. However, the (renormalized) penetration length $m_A^{-1}$ should not constitute an independent diverging length scale. We will see below that this is indeed the case in the dual theory. A last point to note is that the bare coupling $g_0$ also tends to zero when $T \uparrow T_c$. That is, at the tree level the vector field decouples from the theory in this limit and we are left with a pure $|\psi|^4$ theory. As usual, we compute the critical exponents in the symmetric phase of the model, which in the present context corresponds to the superconducting phase.

To facilitate the study of the model we follow Ref. [3] and perform the transformation

$$\psi \to \psi \exp\left(ig_0 \frac{\partial_i}{\nabla^2} h_{0,i}\right) \qquad (4)$$

in order to decouple the longitudinal part of the vector field from the field $\psi$ describing the flux loops. This part is irrelevant for the critical exponents and will be further ignored. We next write the Hamiltonian (2) as a sum of the renormalized Hamiltonian $H$, which is given by (2) without the zeros, and counterterms $\delta H$

$$\delta H = (Z_\psi - 1)|(\partial_i - igh_i)\psi|^2 + (Z_\psi m_0^2 - m^2)|\psi|^2 + u(Z_u - 1)|\psi_0|^4$$
$$+ \frac{1}{2}(Z_h - 1)(\nabla \times \mathbf{h})^2 + \frac{1}{2}(Z_h m_{A,0}^2 - m_A^2)\mathbf{h}^2. \qquad (5)$$

The renormalized objects are related to the bare ones via

$$h_i = Z_h^{-1/2} h_{0,i}, \quad g = Z_h^{1/2} g_0, \quad \psi = Z_\psi^{-1/2} \psi_0, \quad u = Z_u^{-1} Z_\psi^2 u_0. \qquad (6)$$

A few remarks are in order [16]. First, the same Ward identities operate as in the case of a massless vector field. This implies that the minimal coupling to the vector field is preserved when loop corrections are taken into account. We used this observation in writing the relation between the bare and renormalized coupling $g$. Second, the contributions to the self-energy of the vector field are transverse. The mass term of the vector field is consequently not renormalized and does not need a counterterm. In other words, $m_A = Z_h^{1/2} m_{A,0}$. It also follows that the exponent $\gamma_h$ is unaffected by the fluctuations, and retains its mean-field value $\gamma_h = 1$. Incidently, the electric charge does not renormalize in the dual theory since $g = (2\pi/e) m_A$ and both $g$ and $m_A$ renormalize in the same manner.

We now come to an important observation related to the fact that in the dual theory $m_A$ plays the role of a mass as well as of a coupling constant. The standard definition of the critical exponent $\nu$ which determines how the correlation length $m^{-1}$ diverges when the temperature approaches $T_c$: $m^{-1} \sim (T_c - T)^{-\nu}$, viz.,

$$\frac{1}{\nu} = \frac{\partial \ln(m_0^2)}{\partial \ln(m)} \tag{7}$$

can in our case be rewritten as follows

$$\frac{\partial m_{A,0}^2}{\partial \ln(m)} = \frac{m_{A,0}^2}{\nu} \tag{8}$$

because $m_0^2 \sim m_{A,0}^2$ near $T_c$. We use this to cast the $\beta$-function, which is defined by the equation

$$\beta(\hat{g}^2) := m \frac{\partial}{\partial m} \frac{g^2}{m} \bigg|_{u_0, g_0}, \tag{9}$$

with the properly scaled coupling constant $\hat{g}^2 := g^2/m$, in the form

$$\beta(\hat{g}^2) = \hat{g}^2 \left( -1 + \frac{1}{\nu} + \gamma_h(\hat{g}^2, \hat{u}) \frac{\partial \ln(m_A)}{\partial \ln(m)} \right). \tag{10}$$

Here, $\gamma_h(\hat{g}^2, \hat{u})$, with $\hat{u} := u/m$, is the function

$$\gamma_h(\hat{g}^2, \hat{u}) := m_A \frac{\partial}{\partial m_A} \ln(Z_h) \big|_{u_0, g_0} \tag{11}$$

which yields the critical exponent $\eta_h$ when evaluated at the critical point. Without the explicit mass dependence, the coefficient of the $\hat{g}^2$-term in the $\beta(\hat{g}^2)$-function would be $-1$, implying that the origin is an ultraviolet stable fixed point. In (10), however, the coefficient is $-1 + 1/\nu$ which is positive if $\nu < 1$. In this case, the origin becomes infrared stable and the coupled theory reduces to a pure $|\psi|^4$-theory. The best estimate for $\nu$ available from summed perturbation theory at fixed $D = 3$ [15] gives $\nu \approx .6695$, which is smaller than one. Hence, the trivial fixed point $\hat{g}^{*2} = 0$ is infrared stable. This situation differs dramatically from that in the Ginzburg-Landau theory where the coupling $e$ to the vector field has an infrared stable fixed point away from the origin, and the corresponding value $\hat{e}^{*2}$ is too large to allow the coupled system to develop an infrared stable fixed point.

To recapitulate, the dual theory of the superconducting phase transition possesses an infrared stable fixed point given by $\hat{g}^{*2} = 0$ and $\hat{u}^* = \hat{u}^*_{\text{WF}}$, where $\hat{u}^*_{\text{WF}}$ is the Wilson-Fisher fixed point of a pure $|\psi|^4$ theory with reversed temperature axis. The critical exponents of the $\psi$-field are the ones of a superfluid. Since $m$ represents the energy per unit length of an Abrikosov flux tube, the exponent $\nu$ indicates how the tension of these tubes vanishes when the critical point is approached. The critical exponents pertaining to the **h**-field, which physically represents the fluctuating local magnetic field, have their mean-field values. In particular, $\nu_h = 1/2$. Since $m_A$ is the photon mass, this exponent reveals that the magnetic penetration depth diverges near $T_c$ as $(T_c - T)^{-1/2}$, meaning that inside the critical region the empirical formula $m_A^{-1} \sim [1 - (T/T_c)^4]^{-1/2}$ found outside this region remains unchanged.

A last point of interest is the gaussian fixed point, corresponding to $\hat{g}^{*2} = 0, \hat{u}^* = 0$. This fixed point is infrared stable in the $\hat{g}^2$-direction, and unstable in the $\hat{u}$-direction. It describes a theory of vortex loops with a vanishing quartic coupling. In the Ginzburg-Landau picture, this amounts to the situation where the electromagnetic repulsion between two flux tubes precisely balances the attraction mediated by the Higgs particle. This happens when the phase transition changes from second to first order, i.e., at the tricritical point [2]. At the level we are working, the critical exponents characterizing the tricritical point are gaussian. A $|\psi|^6$-term which should be included now will generate logarithmic corrections.

We hope that this Letter initiates experimental effort to study the critical behavior of superconductors. For conventional superconductors the critical region is too small to be probed experimentally. In high-$T_c$ materials, on the other hand, the critical region $|T - T_c|/T_c \sim 10^{-2}$ is large enough to make this study feasible.